\documentclass[a4paper]{elsarticle}
\usepackage{amsmath} 
\usepackage{amssymb}
\usepackage{url}
\usepackage{hyperref}
\usepackage{float}
\usepackage{listings} 
\usepackage[export]{adjustbox}
\usepackage{color}
\definecolor{dkgreen}{rgb}{0,0.6,0}
\definecolor{gray}{rgb}{0.5,0.5,0.5}
\definecolor{mauve}{rgb}{0.58,0,0.82}
\def\rt{R\tau}
\def\rt1{R_1\tau_1}
\def\rt2{R_2\tau_2}
\def\rt3{R_3\tau_3}
\def\rt4{R_4\tau_4}
\def\la{\lambda}

\def\om{\omega}

\def\la{\lambda}
\def\ga{\gamma}
\def\Ga{\Gamma}
\def\de{\delta} 
\def\al{\alpha}
\def\be{\beta}

\def\ep{\epsilon}
\def\si{\sigma}

\def\V{\mathcal V}

\newcounter{bla}

\begin{document}



\title{ALATDYN: a set of Anharmonic LATtice DYNamics codes to compute thermodynamic and thermal transport properties of crystalline solids}      
\tnotemark[1]

\tnotetext[1]{This  research
   project is funded by the National Science Foundation, NSF-CSSI, Office of Advanced Cyberinfrastructure, award number 2103989.}

%

\affiliation[mse]{
    organization={Dept. of Materials Science and Engineering},
    addressline={University of Virginia}, 
    city={Charlottesville},
    postcode={22904}, 
    state={Virginia},
    country={USA}}

\affiliation[phy]{
    organization={Dept. of Physics},
    addressline={University of Virginia}, 
    city={Charlottesville},
    postcode={22904}, 
    state={Virginia},
    country={USA}}

\affiliation[byu]{
    organization={Dept. of Physics and Astronomy}, 
    addressline={Brigham Young University}, 
    city={Provo},
    postcode={84602}, 
    state={Utah},
    country={USA}}
    
\affiliation[mae]{
    organization={Dept. of Mechanical and Aerospace Engineering},
    addressline={University of Virginia}, 
    city={Charlottesville},
    postcode={22904}, 
    state={Virginia},
    country={USA}}
    
\affiliation[msp]{
    organization={Dept. of Mechanical Engineering and Materials Science},
    addressline={University of Pittsburgh}, 
    city={Pittsburgh},
    postcode={15261}, 
    state={Pennsylvania},
    country={USA}}

\affiliation[ut]{
    organization={Department of Mechanical Engineering},
    addressline={University of Tokyo}, 
    city={7-3-1 Hongo, Bunkyo, Tokyo},
    postcode={113-8656}, 
    country={Japan}}

\affiliation[iei]{
    organization={Institute of Engineering Innovation},
    addressline={University of Tokyo}, 
    city={2-11-16 Yayoi, Bunkyo, Tokyo},
    postcode={113-0032}, 
    country={Japan}}
    
\affiliation[sb]{
    organization={Department of Mechanical Engineering},
    addressline={University of California}, 
    city={Santa Barbara},
    postcode={93106}, 
    state={California},
    country={USA}}
    
\affiliation[sun]{
    organization={Research Computing},
    addressline={University of Virginia}, 
    city={Charlottesville},
    postcode={22903}, 
    state={Virginia},
    country={USA}}

\author[mse,phy]{Keivan Esfarjani \fnref{ke}} 

\author[byu]{Harold Stokes}
\author[mae]{Safoura Nayeb Sadeghi}
\author[phy]{Yuan Liang}
\author[mae]{Bikash Timalsina}
\author[ut,iei]{Han Meng}
\author[ut,iei]{Junichiro Shiomi}
\author[sb]{Bolin Liao}
\author[sun]{Ruoshi Sun}

\fntext[ke]{Corresponding author, k1@virginia.edu, orcid=0000-0003-1969-0956}

\hyphenation{DFT
supercell
anharmonic
anharmonicity
Gruneisen
thermophysical}

\begin{abstract}
  We introduce a lattice dynamics package which calculates elastic, thermodynamic  and thermal transport properties of crystalline materials from data on their force and potential energy as a function of atomic positions. The data can come from density functional theory (DFT) calculations or classical molecular dynamics runs performed in a supercell. First, the model potential parameters, which are anharmonic force constants are extracted from the latter runs. Then, once the anharmonic model is defined, thermal conductivity and equilibrium properties at finite temperatures can be computed using lattice dynamics, Boltzmann transport theories, and a variational principle respectively.
  In addition, the software calculates the mechanical properties such as elastic tensor, Gruneisen parameters and the thermal expansion coefficient within the quasi-harmonic approximation (QHA). Phonons, elastic constants and thermodynamic properties results applied to the germanium crystal will be illustrated. Using the force constants as a force field, one may also perform molecular dynamics (MD) simulations in order to investigate the combined effects of anharmonicity and defect scattering beyond perturbation theory.

\vspace{0.5 cm}

\noindent \textbf{PROGRAM SUMMARY}

\begin{small}
\noindent
{\em Program Title: ALATDYN}                                          \\
{\em CPC Library link to program files:} (to be added by Technical Editor) \\
{\em Developer's repository link:} \\
https://github.com/KeivanS/Anharmonic-lattice-dynamics/tree/main \\
{\em Licensing provisions:} GPLv3 \\
{\em Programming language:} FORTRAN90 \\
{\em Nature of problem:}\\
  The ALATDYN suite of codes develops a lattice dynamical model of a crystalline solid. The FOCEX code extracts the model parameters from supercell calculations data on forces versus position and calculates the phonon spectrum, elastic constants and thermodynamic properties within the quasi-harmonic approximation. The SCOP8 code goes beyond QHA and implements the self-consistent phonon theory to minimize the free energy with respect to the strain tensor, atomic positions and harmonic force constants, and thus obtains the state of equilibrium at the given temperature along with the effective phonon bands structure. The THERMACOND code uses the cubic force constants and the crystal symmetries to solve the phonon Boltzmann equation (PhBE) efficiently and deduce the thermal conductivity. Finally, ANFOMOD uses the extracted force constants to perform a molecular dynamics simulation in a supercell. \\
  {\em Solution method:}\\
  Force constants are obtained from a singular value decomposition (or the ridge regression) method. PhBE   is solved by first setting up the collision matrix and effectively inverting it using the conjugate-gradients method. \\
  {\em Additional comments including restrictions and unusual features:}\\
 This code has the ability to extract force constants up to 8th order provided enough force-displacement data is provided. SCOP8 can thus use them to potentially predict phase changes. Coupling of phonons to magnetic or orbital degrees of freedom is an extension we plan to add to the FOCEX and SCOP8 codes in the future. 
   \\

   \end{small}
\end{abstract}


\begin{highlights}
    \item Can extract harmonic and anharmonic force constants up to 8$^{\rm th}$ order, imposing different invariance constraints. 
    \item The state of thermal equilibrium (atomic positions and crystal shape and size) as well as the effective harmonic force constants can be predicted at any temperature and volume (or pressure)
    \item Thermophysical properties such as elastic constants, Gruneisen parameters, free energy, thermal expansion and thermal conductivity from Boltzmann equation (BE) can be predicted  from the knowledge of the extracted force constants
    \item Can also perform molecular dynamics simulations and calculate atomic trajectories and heat current using the extracted anharmonic force field. This is the more accurate alternative to solving BE, as it treats anharmonicity and mass disorder beyond perturbation theory.
\end{highlights}


\maketitle

\section{Introduction}

The theory of lattice dynamics, based essentially on the physics of the harmonic oscillator has shaped the science behind the thermophysical properties of materials. Nowadays, with the advances in computer software and hardware, calculations of the electronic ground state properties of materials, based on the density functional theory has become routine. Total energy and forces on atoms are an output of such theories, and this has allowed us \cite{keivanprb2008} to develop methodologies to compute force derivatives, or force constants (FCs), in order to obtain the parameters of an anharmonic lattice dynamics model. Many softwares are nowadays available to compute force constants of crystalline solids. Among them, we can name ALAMODE \cite{alamode,Tadano2018}, ShengBTE \cite{ShengBTE2014}, AlmaBTE \cite{almabte}, Phonopy \cite{togo2015}, SSCHA \cite{Monacelli2021} , Hiphive \cite{Erhart2019}, TDEP \cite{Hellman2013-1,Hellman2013-2}, Phonon \cite{parlinski1997}, Phon \cite{alfe2009} ... and many others. Here, we are presenting a new set of codes based on our original paper \cite{keivanprb2008} to compute thermophysical properties of anharmonic crystals.
These codes are similar in capabilities but  different in the way the FCs are extracted and processed. Our code to calculate the thermal conductivity (THERMACOND) uses full symmetry of the crystal to compute the non-equilibrium distribution function more efficiently. Codes to compute the free energy and the state of equilibrium at high temperatures, as well as the computation of mechanical properties are relatively novel\cite{Nabr,Diaz}. In addition, we provide codes (ANFOMOD) to perform molecular dynamics simulations and calculate the heat current with the extracted polynomial force field. 
In what follows, we describe the methodology and capabilities of the code to extract force constants, FOCEX, which stands for FOrce Constants EXtraction. The FCs are extracted from a set of displacement-force data on atoms in one or many supercells. They can be used as input to the code ANFOMOD (ANharmonic FOrce MOlecular Dynamics) to perform MD simulations and calculate the heat current and thermal conductivity from the exact Green-Kubo formula, thereby incorporating anharmonicity and defect scattering beyond perturbation theory.
Our other codes, which take as input the extracted force constants, compute the thermal conductivity (THERMACOND) and the state of thermal equilibrium, including structural phase changes, at finite temperatures (SCOP8). They will be described in other upcoming papers.

\section{The FOrce Constants EXtraction (FOCEX) code}

Let us start by defining the anharmonic lattice dynamics (LD) model potential energy $V^{LD}$, which is a Taylor expansion of the potential energy of a crystal in powers of atomic displacements measured from a reference (usually a lattice-) position.
The model potential energy can be written in the following form:

\begin{equation}
    V^{LD} = E_0 + \sum_i \Pi_i u_i +  \frac{1}{2!} \sum_{ij}\Phi_{ij} \, u_i u_j +  \frac{1}{3!} \sum_{ijk}\Psi_{ijk} \,
    u_i u_j u_k + \frac{1}{4!} \sum_{ijkl}\chi_{ijkl} \, u_i u_j u_k u_l+\cdots
    \label{actualPotential}
\end{equation}

where the roman index $i$ labels the triplet $(R,\tau,\alpha)$ with $R$
being a translation vector of the primitive lattice, $\tau$  refers to an atom within the
primitive unit cell, and $\alpha$ is the Cartesian coordinate of the atomic
displacement $u$. In other words,  $u_i$ is the displacement of the
atom $(R,\tau)$ in the direction $\alpha$ from its 
reference position defined by the vector  $(R+\tau)$. $\Phi,\Psi$ and $\chi$ are respectively the harmonic, cubic and
quartic FCs, whereas $\Pi$ is the negative of the residual force on atom $i$,
and is zero if the potential energy $V$ is expanded around its minimum
or ground state configuration where  $(R,\tau)$ are the equilibrium positions of the atoms. Although this is usually the case, i.e. $\Pi_i=0$, in low-symmetry crystals, this term maybe non-zero but small due to imperfect convergence in atomic relaxations.  In clusters or molecules the formalism is the same, only the
translation vector $R$ needs to be dropped. The resulting LD model force on atom $i$ would then be

\begin{equation}
    F_i^{LD} = - \frac{\partial V}{\partial u_i} = -\Pi_i -\sum_j  \Phi_{ij} \,u_j - \frac{1}{2} \sum_{jk}\Psi_{ijk} \,
    u_j u_k - \frac{1}{3!} \sum_{jkl} \chi_{ijkl} \,  u_j u_k u_l+\cdots
    \label{forcemodel}
\end{equation}

Although this expansion has been truncated to order 4, the code FOCEX can handle expansions to arbitrarily high orders provided there is enough computer memory to handle the higher-rank force constants. 
Besides this truncation in rank, the other approximation controlled by the user is in the truncation of the {\bf range of interactions}.
Furthermore, the FCs of the system are reduced based on permutation, translation, rotation and point or space group symmetries of the crystal.  Hence, the FOCEX code first starts by identifying the  symmetry properties and constructing the space group matrices $S$ and translations, in order to define the set of  irreducible or independent force constants, and their relationship to the full set of force constants. 

\subsection{Invariance under permutation of indices} 
From the above expression[\ref{forcemodel}], it is evident that the FCs are derivatives of potential energy, so the order of derivatives does not matter and thus the indices of FCs can be arbitrarily swapped: 

\begin{equation}
    \Phi_{ij} = \Phi_{ji} \\
    \Psi_{ijk} = \Psi_{ikj} = \Psi_{jik} = \Psi_{kji} = ... \\
    \chi_{ijkl} = \chi_{ikjl}=\chi_{ijlk}=\chi_{jikl} = ...
    \label{esfarjani2008method3}
\end{equation}

\subsection{Invariance under arbitrary translations and rotations}
The potential energy of a solid remains the same under arbitrary translations and rotations of that solid. For any given atomic displacement $u$, the potential energy associated with the translation of the whole system by an arbitrary vector $c$ is $V(u+c)$ and it must satisfy $V(u+c)=V(u)$ and $F(u+c)=F(u)$, where $u$ are the dynamical variables and $c$ is the arbitrary constant shift vector. Identifying terms order by order we arrive at the following equations \cite{liebfried}: 

\begin{equation}
\begin{split}
    0 &= \sum_{\tau} \, \Pi_{0 \tau \alpha} \,\,\, \forall (\alpha) \,\,  ({\rm Total \,~force \,~on\,~unit\,~cell = 0})  \\
    0 &= \sum_{R_1,\tau_1} \, \Phi_{0\tau,R_1 \tau_1}^{\alpha\beta} \,\,\, \forall (\alpha\beta,\tau) \\
    0 &= \sum_{R_2,\tau_2} \, \Psi_{0\tau,R_1\tau_1,R_2\tau_2}^{\alpha\beta\gamma} \,\,\, \forall(\alpha\beta\gamma,\tau,R_1\tau_1) \\
    0 &= \sum_{R_3,\tau_3} \, \chi_{0\tau,R_1\tau_1,R_2\tau_2,R_3\tau_3}^{\alpha\beta\gamma\delta} \,\,\,  \forall(\alpha\beta\gamma\delta,\tau,R_1\tau_1,R_2\tau_2)
\end{split}
\label{asr}
\end{equation}

These are the well-known {\it acoustic sum rules} generalized to higher-order FCs.

Similarly, the force on atoms and the total potential energy of the crystal  should not change under arbitrary rotation of the crystal. This  leads to the following constraints on FCs \cite{liebfried}:

\begin{equation}
    \begin{split}
        0 &= \sum_{\tau} \,  \Pi_{0 \tau\alpha}\, \tau^{\beta}\, \varepsilon^{\alpha\beta\nu}, \,\,\, \forall ( \nu) \,\, ({\rm Torque \,on\,unit\,cell}=0) \\
    0 &=  \sum_{R_1,\tau_1} \, \Phi_{0\tau,R_1 \tau_1}^{\alpha\beta}\, (R_1+\tau_1)^{\gamma}\, \varepsilon^{\beta\gamma\nu}
     + \Pi_{0 \tau}^{ \beta }\, \varepsilon^{\beta \alpha \nu}  \,\,\, \forall (\alpha\nu,\tau) \\
    0 &= \sum_{R_2,\tau_2} \, \Psi_{0\tau,R_1\tau_1,R_2\tau_2}^{\alpha\beta\gamma} \, (R_2+\tau_2)^{\delta} \, \varepsilon^{\gamma\delta\nu} + \Phi_{0\tau,R_1\tau_1}^{\gamma\beta}\, \varepsilon^{\gamma\alpha\nu}  + \Phi_{0\tau,R_1\tau_1}^{\alpha\gamma}\, \varepsilon^{\gamma\beta\nu} \,\,\, \forall(\alpha\beta\nu,\tau,R_1\tau_1) \\
    0 &= \sum_{R_3,\tau_3}\, \chi_{0\tau,R_1\tau_1,R_2\tau_2,R_3\tau_3}^{\alpha\beta\gamma\delta}\,
    (R_3+\tau_3)^{\mu}\, \varepsilon^{\delta\mu\nu}
    + \, \Psi_{0\tau,R_1\tau_1,R_2\tau_2}^{\delta\beta\gamma} \,\varepsilon^{\delta\alpha\nu} +
     \, \Psi_{0\tau,R_1\tau_1,R_2\tau_2}^{\alpha\delta\gamma} \,\varepsilon^{\delta\beta\nu}   \\
    & \quad + \, \Psi_{0\tau,R_1\tau_1,R_2\tau_2}^{\alpha\beta\delta} \,\varepsilon^{\delta\gamma\nu}
    \,\, \, \forall(\alpha\beta\gamma\nu,\tau,R_1\tau_1,R_2\tau_2)
    \end{split}
    \label{rotational}
\end{equation}

where $\varepsilon^{\alpha\beta\gamma}$ is the anti-symmetric Levi-Civita symbol.
Moreover, an implicit
summation over repeated Cartesian indices is implied.

We see that rotational invariance relations relate the first to the second-order terms, the second to the third-order terms and the third to the fourth-order terms, respectively. This implies that if the expansion of the potential energy is truncated after the fourth order terms, we must start with the fourth order terms, and the application of the rotational invariance rules will give us constraints on the third and lower-order FCs. For more details on the proof regarding the translation and rotational invariance relations, we refer the reader to the article by Liebfried and Ludwig \cite{liebfried}. 

These invariances are usually not imposed in other methods, but become important in 2D materials to correctly describe the out-of-plane ZA modes and in 1D materials to describe the bending modes, both of which must have a quadratic dispersion at small phonon wavenumbers. For any simplified LD model to physically make sense, these invariances must be imposed.

In addition to these symmetries, it is important to note that elastic constants are subject to a number of constraints that enforces them to be symmetric under the exchange of Voigt indices. Huang discusses these invariances\cite{bornhuang}, which provide an additional 15 set of constraint equations for low symmetry crystals with internal degrees of freedom in the primitive cell \cite{sluiter1998elastic,sluiter1999fcalloys}].
These relations are explicitly written in section \ref{elasticity} where we define the elastic constants.

\subsection{Invariance under point or space group symmetry}

Other symmetry operations such as rotations or mirror symmetries, elements of the space group  also leave the potential energy of the crystal unaltered \cite{Maradudin1962}. In particular, force constants must be invariant under translations of lattice vectors, implying the following relations: 

\begin{equation}
    \begin{split}
        \Pi_{R\tau\alpha} &= \Pi_{0 \tau\alpha} \,\, \forall (R \tau \alpha)  \\
    \Phi_{R\tau,R_1\tau_1}^{\alpha\beta} &= \Phi_{0 \tau ,(R_1-R) \tau_1}^{\alpha\beta}  \\
    \Psi_{R\tau,R_1\tau_1,R_2\tau_2}^{\alpha\beta\gamma} &= \Psi_{0 \tau ,(R_1-R) \tau_1,(R_2-R) \tau_2}^{\alpha\beta\gamma}  \\
    \chi_{R\tau,R_1\tau_1,R_2\tau_2,R_3\tau_3}^{\alpha\beta\gamma\delta} &= \chi_{0 \tau ,(R_1-R) \tau_1,(R_2-R) \tau_2, (R_3-R) \tau_3}^{\alpha\beta\gamma\delta}
    \end{split}
    \label{esfarjani2008method6}
\end{equation}

These equations are used to restrict all the force constants to those whose first index is an atom in the central primitive cell ($R=0$). This is how they are stored in the FOCEX code.

If the point or space group related operation $S$  is governed by the  matrix $S$ then the following relations must hold:

\begin{equation}
    \begin{split}
        \Pi_{S\tau\alpha} &= \sum_{\alpha'} \,\Pi_{\tau\alpha'} \, {\cal S}_{\alpha,\alpha'} \\
    \Phi_{ S\tau, S\tau_1}^{\alpha\beta} &= \sum_{\alpha'\beta'} \,\Phi_{\tau,\tau_1}^{\alpha'
    \beta'} \, {\cal S}_{\alpha,\alpha'}\, {\cal S}_{\beta,\beta' }  \\
    \Psi_{ S\tau, S\tau_1, S\tau_2}^{\alpha\beta\gamma} &= \sum_{\alpha'\beta'\gamma'} \,\Psi_{\tau,\tau_1\tau_2}^{\alpha' \beta'\gamma'} \, {\cal S}_{\alpha,\alpha'}\, {\cal S}_{\beta,\beta' } \, {\cal S}_{\gamma,\gamma' }   \\
    \chi_{ S\tau, S\tau_1, S\tau_2, S\tau_3}^{\alpha\beta\gamma\delta} &= \sum_{\alpha'\beta'\gamma'\delta'} \,\chi_{\tau,\tau_1,\tau_2,\tau_3}^{\alpha' \beta' \gamma' \delta'} \, {\cal S}_{\alpha,\alpha'}\, {\cal S}_{\beta,\beta' } \, {\cal S}_{\gamma,\gamma'}\, {\cal S}_{\delta,\delta' }
    \end{split}
    \label{space_group}
\end{equation}

where ${\cal S}_{\alpha,\alpha'}$ are the $3\times3$ matrix elements of the symmetry operation $S$ , and $S \tau$ is the image of the atom $\tau$ under that space group transformation (including eventual glide operations in non-symmorphic groups). 

All these symmetry relations impose a {\it linear} set of constraints on the force constants. 

These invariance relations must be satisfied by any physically meaningful description of force constants. On the other hand, we approximate the Taylor expansion of the potential energy by truncating the range of FCs and their order. As a result, applying the constraints may slightly shift their value from its real numerical value, but it has the advantage that they are physical and, for instance, will replicate the linear acoustic phonon dispersion or quadratic out of plane in 2D or bending modes in 1D at $k=0$. Therefore, one should be aware that an unrealistic truncation to a too small range may result in disagreement with experimental values. Therefore checks must be performed, and range increased until the results on phonon dispersions for instance do not change above some cut-off value.

\subsection{Extraction procedure}

To extract force constants of the LD model, we perform accurate calculations of forces on displaced atoms in a supercell, typically using DFT methods, and fit their values with the polynomial LD model described in Eq. (\ref{forcemodel}). 
The set of invariance relations \ref{asr},\ref{rotational} and \ref{space_group} form a (non-square) matrix relation of the form $[ I ] \Phi = 0$, while the force-displacement constraints is of the form $[ U ] \Phi = F $.
These two parts are combined together, and 
a singular value decomposition (SVD) or rather a ridge regression algorithm is used to extract the independent FCs. The reason is that the above set of equations is usually an 
overcomplete set. 
Another option available to the user is to first fit (using SVD) true forces to Eq. (\ref{forcemodel}), i.e. solve $[ U ] \Phi = F $ in a least squares sense, and then to project on the zero eigenvalue space of the $[ I ]$ matrix. This enforces the invariance relations (\ref{asr}), (\ref{rotational}) and (\ref{space_group})  more exactly, but at the cost of violating the $[ U ] \Phi = F $ relation.

It is to be noted that the data set obtained for force-displacement in fitting comes from a supercell. The size of supercell should exceed the range of FCs or else, the contribution of the images of atoms will also be included in the computed FCs: meaning for instance only the sum $\sum_L  \Phi_{\tau,L+\tau'}^{\alpha\beta}$, where $L$ is a supercell translation vector and $\tau$ and $\tau'$ are two atoms within the supercell, is accessible.

Additionally, FOCEX has the ability to read input force constants (typically harmonic) from a previous fit, and extract only the remaining non-fitted (typically cubic and quartic) FCs.

\subsection{Phonon dispersion and their density of states}

With the extracted force constants, and since the calculation does not take much time and resources, we have added to FOCEX the capability to calculate the phonon dispersion, density of states (DOS), mode Gruneisen parameters, the elastic tensor, and the thermodynamic properties within the quasi-harmonic approximation, which are all relatively fast calculations even on a laptop computer.
We use the following definition for the $\al \be$ components of dynamical matrix at the wavenumber $k$ 
\begin{equation}
    D_{\tau,\tau'}^{\alpha \beta}(k) = \sum_R \frac{\Phi_{\tau,R+\tau'}^{\alpha\beta} }{\sqrt{ m_{\tau} m_{\tau'}}} \, e^{ik.(R+\tau'-\tau)} 
    \label{dynmat}
\end{equation}
This is a Hermitian matrix, the eigenvalues of which are the square of phonon frequencies at $k$, also satisfying completeness and orthonormality.
\begin{equation}
      D_{\tau,\tau'}^{\alpha \beta} (k) \,e_{\tau' \beta,\lambda} (k)=\omega_{\lambda} ^2 (k) \,e_{\tau \alpha,\lambda}(k) 
\end{equation}

 \begin{equation}
    \sum_{\la} e_{\tau \alpha,\lambda}(k)^* e_{\tau' \be,\lambda}(k) = \de_{\al\be} \de_{\tau \tau'}  \,\,\,\, and \,\,\,\, \\
   \sum_{\tau \al} e_{\tau \alpha,\lambda}(k)^* e_{\tau \alpha,\la'}(k) = \de_{\la \la'}  
 \end{equation}

In other words, if one considers ${\tau \alpha}$ as the line index and ${\lambda}$ as the column index of the eigenvector matrix, this matrix should be unitary.
There is another choice for the phase of the dynamical matrix, which is to replace the factor $e^{ik.(R+\tau'-\tau)}$ in eq. \ref{dynmat} by $e^{ik.R}$.
Both choices lead to the same phonon frequencies but the eigenvectors in the second case will be $e_{\tau \alpha,\lambda}(k) e^{-ik.\tau} $. However, the first choice leads to a continuous eigenvector as a function of $k$ when the Brillouin zone is crossed.
This choice of the definition of the dynamical matrix leads to well-defined group velocities, and that is what is implemented in FOCEX. The expression for the group velocity is obtained by taking a wavevector derivative of the square of the phonon frequency ($\om^2=e.D.e \Rightarrow 2 \om v= e.(dD/dk).e$). Using the Hellmann-Feynman theorem applied to the derivative of the dynamical matrix, and dividing by $2 \om_{\la}(k)$ one finds:

\begin{equation}
   v_{\la}^{\ga} (k) =\frac{1}{2 \om_{\la}(k)}  \sum_{\tau\al,R\tau' \al'} i (R+\tau'-\tau)^{\ga} \, e_{\tau \alpha,\lambda}(k)^* 
   \frac{\Phi_{\tau,R+\tau'}^{\alpha\beta} }{\sqrt{ m_{\tau} m_{\tau'}}} 
   e_{\tau \alpha,\lambda}(k) \, e^{ik.(R+\tau'-\tau)}
\label{velocity}
\end{equation}

In case of degenerate bands, a rotation of the eigenvectors of the degenerate subspace will be needed to make $dD/dk$ diagonal, leading to the correct group velocities on the diagonal. 

The phonon density of states is calculated using both the tetrahedron method in the full first Brillouin zone (FBZ) and the Gaussian-smearing method in the irreducible Brillouin zone (IBZ). The latter will be inaccurate unless a large number of k-points are used in the sampling of the IBZ, while the former is almost converged for a moderate grid size on the order of 20 k-mesh per direction.

\subsection{Non-analytic correction for ionic dielectrics} 
For such systems, the input to the file \texttt{dielectric.params} which contains the electronic contribution to the dielectric matrix $\epsilon_{\infty}$ and the Born effective charges $Z_{\tau}^{\al\be} $ of every atom $\tau$ in the primitive cell, must be properly initialized. These numbers are obtained from a separate DFPT calculation\cite{Gonze1997,Baroni2001}. 
The contribution of the ionic charges on the force constants is long-ranged and cannot be captured by the small supercell. It is usually calculated using the Ewald sum method \cite{ewald,Ohno1999,Gonze1997}, which is formed from the the sum of short-range real space component and a long-range reciprocal-space component. In FOCEX, once the harmonic force constants  are fitted, we calculate the dynamical matrix on a grid of reciprocal lattice vectors of the supercell inside the Wigner-Seitz cell of the primitive lattice, including its boundaries, albeit with a weight inversely proportional to the number of images that also fall on the Brillouin zone boundaries. This corresponds to  the Fourier transform of the real-space force constants. Phonon frequencies on this grid, $D(G_{\rm sc})$, are exact. From these dynamical matrices, we subtract the non-analytical contribution\cite{Gonze1997}, and Fourier transform back to real space, in-order to presumably obtain a short-ranged set of force constants, with a range confined to the Wigner-Seitz cell of the supercell. Then at an arbitrary wavevector $q$, we use Fourier interpolation to calculate the dynamical matrix from the latter set of force constants, to which the non-analytical part that was subtracted earlier, is added back (see also \cite{Togo2023}).

\section{Some notes on Elastic constants} \label{elasticity} 
It is possible to extract the elastic constants from the knowledge of the force constants\cite{Wallace1998}. Below, we outline a simpler method and provide the final formulas. 

Under a uniform applied strain, denoted by $\eta_{\al\be}=\partial u_{\al} / \partial r_{\be}$ a point $x$ of the medium is  moved to $x'_{\al}=x_{\al}+\eta_{\al\be}x_{\be}$, and the total energy of the harmonic crystal is increased, by definition of the elastic constants, by $\Delta E/V_0 =  C_{\al\be,\ga\de} \ep_{\al\be} \ep_{\ga\de}/2 $, where $V_0$ is the unit cell volume and the Cauchy strain tensor $\ep$ is defined as $\ep_{\al\be}=(\eta_{\al\be}+\eta_{\be\al})/2$. The latter is used because the antisymmetric part of the strain tensor $\eta$, represents a pure rotation and  does not contribute to the total energy change. So, using the Cauchy strain, the 9 strains are reduced to 6 independent ones.
Under such strain, we can use eq. \ref{actualPotential} to calculate the potential energy increase by replacing $u_i$ by:
\begin{equation}
    u_{R\tau\al}(t)=\eta_{\al\be} (R+\tau)^{\be} +u_{\tau\al}^0 + y_{R\tau\al}(t)=S_{R\tau\al}+y_{R\tau\al}(t)
\end{equation} where $S$ is the static displacement, containing an extra-term $u^0$ which represents the relaxation of atom $\tau$ within the primitive cell in addition to that due to the uniform deformation dictated by $\eta$. This can occur in low-symmetry crystals, where atoms may not completely follow the uniform deformation, and require an extra displacement $u_{\tau\al}^0$ to minimize the potential energy. The last term $y_{R\tau\al}(t)$ represents the dynamical motion of the atom $R\tau$, or phonon degree of freedom,  which has a time average of zero by construction. Plugging into eq.\ref{actualPotential}, we can derive the effective harmonic potential, at finite temperatures, as the coefficient of the $y^2$ term, and the elastic tensor as the second derivative of 
the average potential energy with respect to the applied (Cauchy) strain: $C_{\al\be,\ga\de} =\frac{1}{V_0}\frac{\partial^2 E}{\partial \ep_{\al\be} \partial \ep_{\ga\de}} $.
Note that as a second derivative, $C$ must be symmetric under swapping of the order of differentiation $\al\be \longleftrightarrow \ga\de$. Due to the symmetry of the Cauchy strain tensor $\ep$ itself, the elastic tensor $C$ will also be invariant with respect to swap of $\al \longleftrightarrow \be$ and $\ga \longleftrightarrow \de$ separately.

Substitution of deformations $S$ in eq.\ref{actualPotential}, keeping up to harmonic terms in $S$, 
leads to:
\begin{equation}
    (V^{LD}-E_0)(S)=\sum_{R\tau\al} \Pi_{\tau\al} S_{R\tau\al}+\frac{1}{2} \sum_{R\tau\al,R'\tau'\al'} \Phi_{R\tau,R'\tau'}^{\al\al'} \, S_{R\tau\al}S_{R'\tau'\al'}+... 
    \label{vofetau0}
\end{equation}

While this is valid strictly at zero temperature, this discussion can be extended to finite temperatures using the self-consistent phonon theory\cite{Werthamer1970} where the harmonic force constant $\Phi$ is replaced by an effective temperature-dependent one\cite{esfarjani2020,esfarjani2022}: \begin{equation}
    \Phi \rightarrow K(T)=\Phi+\Psi S +\chi (S S + \langle y y \rangle )/2
    \label{keff}
\end{equation}. 
In what follows, we will derive the (harmonic) elastic properties, strictly speaking, at zero temperature, but the above formula may be used as an extension to finite temperatures, by replacing every occurrence of $\Phi$ in the following by $K(T)$.

The stress $\si$ is obtained by taking the first derivative of the energy in eq. \ref{vofetau0} with respect to the Cauchy strain $\ep$, while the second derivative  gives the elastic tensor.

Compared to the method discussed by Wallace \cite{Wallace1998}, here, we are proposing an equivalent but simpler method to obtain elastic constants. We first write the harmonic potential energy in eq. \ref{vofetau0}, phenomenologically, up to second powers of internal coordinates  $u^0$ and strain $\eta$ as: 
\begin{multline}
    V^{LD}(\eta,u^0) = E_0+\Pi_{\tau\al} u^0_{\tau\al} +  \Xi_{\mu\nu}\eta_{\mu\nu} + \\
        \frac{1}{2} \left[ \phi_{\tau,\tau'}^{\al\al'} \,u^0_{\tau \al}u^0_{\tau' \al'}
+ A_{\al\be,\al'\be'} \, \eta_{\al\be}\eta_{\al'\be'} + 2 Q_{\tau\al;\al'\be'} \,u^0_{\tau \al}\eta_{\al'\be'} \right]
\label{V_eff}
\end{multline}
As before, sums over repeated indices are implied.  
By comparing this equation with eq.\ref{actualPotential} in which $u_{R\tau\al}$ is replaced  by $\eta_{\al\be}(R+\tau)^{\be} + u^0_{\tau\al}$, 
the new coefficients $A,Q,\phi$ can be identified with the harmonic force constants $\Phi$ through the following relations:
\begin{eqnarray}
\Xi_{\al\be} &=& \frac{1}{N}\sum_{R\tau} \Pi_{\tau \al} (R+\tau)^{\be}=\sum_{\tau} \Pi_{\tau \al} \tau^{\be} \nonumber \\
\phi_{\tau,\tau'}^{\al\al'} &=&\frac{1}{N}\sum_{R,R'} \Phi_{R\tau,R'\tau'}^{\al,\al'} \,\,\,\, {\rm with} \,\,\,\, \phi_{\tau,\tau'}^{\al\al'} =\phi_{\tau',\tau}^{\al'\al}  \nonumber \\
    Q_{\tau\al;\al'\be'}&=&\frac{1}{N}\sum_{R R'\tau'} \Phi_{R\tau,R'\tau'}^{\al,\al'} (R'+\tau')^{\be'}=\frac{1}{N}\sum_{R R'\tau'} \Phi_{R\tau,R'\tau'}^{\al,\al'} (R'+\tau'-R-\tau)^{\be'}   \nonumber   \\    
    A_{\al\be,\al'\be'} &=&\frac{1}{N}\sum_{R\tau,R'\tau'} \Phi_{R\tau,R'\tau'}^{\al,\al'} (R+\tau)^{\be} (R'+\tau')^{\be'}  \,\,{\rm with} \,\,A_{\al\be,\al'\be'} =A_{\al'\be',\al\be}  \nonumber \\
     {\rm ASR}&=>& \sum_{\tau} \phi_{\tau,\tau'}^{\al\al'} =\sum_{\tau'} \phi_{\tau,\tau'}^{\al\al'} =0  \nonumber \\
     {\rm ASR}&=>& \sum_{\tau} Q_{\tau\al;\al'\be'}=0  \nonumber 
\end{eqnarray}

In addition to translational invariance (ASR), rotational invariance implies $Q_{\tau\al;\be\ga} \varepsilon^{\al\be\ga}=0 ; \,\, \forall \tau 
$. 
Even though, due to translational symmetry of the crystal, the above sums do not depend on the vectors $R$, we have kept them  in the summation for the sake of symmetry in the formulas,  but divided by $N$, the (infinite) number of translation vectors $R$. If there are $N_0$ sites in the primitive cell, $\Pi$ is a $3N_0$ component array, $\phi$ and $Q$  are respectively a $3N_0\times 3N_0$ and $3N_0\times 6$ matrices, while $A$ is will be a $6\times6$ matrix if the Voigt notation is used to describe the strain and stress tensors. Voigt notation is used after symmetrizing $A$ and $Q$ since the potential energy should not depend on the rotational part of the strain $\eta$.

The key point in  calculating the elastic tensor, which is the second strain-derivative of $V^{LD}$, is to 
 first relax the crystal by minimizing the potential energy with respect to $u^0$. This corresponds to the equilibrium condition: $\frac{\partial V^{LD}}{\partial u^0_{\tau\al}}=0= \Pi+\phi u^0 + Q \eta 
=> u^0(\eta) = -\Ga (\Pi+Q \eta)$
where the matrix $\Ga$ is the inverse of $\phi$: 
$\sum_{\tau',\al'}\Ga_{\tau,\tau'}^{\al\al'} \phi_{\tau',\tau''}^{\al'\al''} = \de_{\al\al''} \de_{\tau\tau''}$. 

Note that since $\phi$ has three acoustic modes of zero eigenvalue, it is not invertible. Fixing the residual  displacement of atom 1 to be $u^0_1=0$ makes $\phi$ an invertible $(3N_0-3)\times(3N_0-3)$ matrix, so that in the sums over primitive cell atoms $\tau$ one should exclude the first atom, or alternatively, one can pad the matrix $\Gamma$ by 0 to make it $3N_0 \times 3N_0$.
\begin{multline}
    u^0_{\tau\al}(\eta)=-\Ga_{\tau,\tau'}^{\al \al'} (\Pi_{\tau'\al'}+Q_{\tau'\al';\mu\nu}\, \eta_{\mu\nu})  \\ {\rm with} \,\,\,\Ga_{1,\tau'}^{\al\al'}= \Ga_{\tau,1}^{\al\al'}=0 \,\,
    {\rm leading \,automatically\, to} \,\, u^0_{1 \al}=0. 
    \label{u0} 
\end{multline}

Note that even in the absence of strain ($\eta=0$),  if there is a residual force $\Pi_{\tau \al}$ in the reference configuration, equation \ref{u0} predicts the  atom $\tau$ should relax to its equilibrium position given by: $u^0_{\tau \al}=-\Ga_{\tau,\tau'}^{\al\al'} \Pi_{\tau' \al'}$.

After relaxing the internal coordinates $u^0$, and substituting the formula in eq. \ref{u0} for $u^0$ in eq. \ref{V_eff}, the potential energy becomes a function of strain $\eta$ only:
\begin{multline}
    V^{LD}_{relaxed}(\eta)=E_0-\frac{1}{2} \Pi_{\tau\al} \Ga_{\tau\al,\tau'\al'}\Pi_{\tau'\al'} + \Xi_{\mu\nu}\eta_{\mu\nu} 
    -\Pi_{\tau\al} \Ga_{\tau\al,\tau'\al'}
Q_{\tau'\al';\mu\nu} \eta_{\mu\nu} \\
    +\frac{1}{2}
( A_{\mu\nu,\mu'\nu'}  - Q_{\tau\al;\mu\nu}\Ga_{\tau\al,\tau'\al'}Q_{\tau'\al';\mu'\nu'} )
 \eta_{\mu\nu}\eta_{\mu'\nu'}   
 \label{vofeta}
\end{multline}
Finally, the first and second strain derivatives  give respectively the residual stress and the elastic tensor. Using the chain rule to convert $\eta$ derivatives to $\ep$ derivatives, essentially boils down to symmetrizing the results (in effect, the symmetric parts of $A$ or $Q$ come in the results as previously mentioned). 
\begin{equation}
    \sigma_{residual}=\frac{d V^{LD}_{relaxed} }{V_0 \, d\eta}=\frac{d V^{LD}_{relaxed} }{V_0\, d\ep} = \frac{1}{V_0} {\rm Symmetrized} \left(\Xi - \Pi \,\Ga \,Q \right)
    \label{residual_stress}
\end{equation}
\begin{equation}
    C=\frac{1}{V_0} \frac{d^2 }{d \ep^2}V^{LD}_{relaxed}(\ep,u_{relaxed}^0(\ep))=\frac{1}{V_0} {\rm Symmetrized}(A-Q\Ga Q)
\end{equation}
where $V_0$ is the equilibrium unit cell volume, and ${\rm Symmetrized}(B_{\mu\nu})=
\frac{1}{2}(B_{\mu\nu}+B_{\nu\mu})$ and  ${\rm Symmetrized}(B_{\mu\nu,\mu'\nu'})=
\frac{1}{4}(B_{\mu\nu,\mu'\nu'}+B_{\nu\mu,\mu'\nu'}+B_{\mu\nu,\nu'\mu'}+B_{\nu\mu,\nu'\mu'})$. 
While in the elastic tensor,  the first term $A$ is the second {\it partial} derivative of the potential energy with respect to strain, the second term $Q\Ga Q$ is the correction due to the atomic relaxations $u^0$, reflecting that the elastic constant is the full second derivative. Note that due to the symmetry of $\phi$ and thus of $\Gamma$, this term is also symmetric with respect to interchange of $\mu\nu \leftrightarrow \mu'\nu'$.

Similar to eq. \ref{u0}, which gives the {\it relaxed positions} of the primitive cell atoms in the reference configuration, one can also, get the final {\it relaxed shape} of the primitive cell in terms of the residual stress in eq. \ref{residual_stress}:
\begin{equation}
    \frac{d V^{LD}_{relaxed} }{V_0 \, d\ep}=0=\sigma_{residual}+C \ep  =>\epsilon^{eq} = -C^{-1}\sigma_{residual}
    \label{ep0}
\end{equation}

Note that even if the reference configuration is not exactly the total energy minimum, due to incomplete or unconverged numerical relaxation,  eqs. \ref{ep0} and \ref{u0} can in principle be applied (in this order as eq. \ref{u0} involves the strain) to further relax the reference primitive cell  (these corrections are non-zero only if $\Pi \ne 0$). This can happen in crystals of low symmetry, where space group symmetry constraints do not imply $\Pi = 0$. In cases where $\Pi=0$ due to symmetry, such as in Ge, there might additionally be a residual stress because the lattice constant relaxation was not complete. This residual stress is however available at the end of the DFT-SCF relaxation and should be added to the expression for $\sigma_{residual}$ if the equilibrium strain in eq. \ref{ep0} is needed.

Also note that since typically force constants are on the order of 10 eV/\AA$^2$, if the residual force is on the order of $\Pi \simeq 10 $ meV/\AA, the corresponding required atomic relaxation would be on the order of $u^0_{\tau} \simeq \Pi_{\tau}/\phi_{\tau,\tau} \simeq $ 0.001 \AA$\,$ and therefore usually small. 

The isotropic and uniaxial elastic constants are denoted respectively by the Bulk and Young moduli $B$ and $Y$. 

\subsection{Bulk modulus}

The isothermal bulk modulus is defined as  $B_T=-V \left(\frac{dP}{dV} \right)_{P=0,T}$. The isotropic compression condition translates into $\si_{xx}=\si_{yy}=\si_{zz}=-P; \, \si_{\al\be}=0 \, {\rm for} \, \al\ne \be$, and since the volume change is given by $\Delta V/V=\ep_{xx}+\ep_{yy}+\ep_{zz}={\rm Tr}\,\ep$, one can express $B$ in terms of elastic or compliance tensor elements as: 
\begin{equation}
    B = \frac{{\rm Tr} \sigma /3}{ {\rm Tr}\,\ep}=\frac{1}{9} {\sum_{i,j=1,2,3} C_{ij}} =\frac{1}{\sum_{i,j=1,2,3} S_{ij}} 
\end{equation}
 where $S$  is the inverse of $C$ when written as a 6$\times$ 6 matrix in the Voigt notation, and is called the {\it compliance tensor}.
For crystals of cubic symmetry, it can be simplified as:
$ (C_{11}+2C_{12})/3 $.  

\subsection{Shear modulus}

The shear stress $\mu_{xy}$ is the ratio $\left( \frac{\partial \si_{4}}{\partial \ep_{4}} \right) _{\si_{j\ne 4}=0} $. 
It can therefore be written as: $\mu=1/S_{44}$. In some works, however, in order to average over anisotropy, e.g. when dealing with polycrystalline samples, where the elastic properties are supposed to be isotropic, usually the formula $\mu_{iso}=(C_{11}-C_{12}+3 C_{44})/5 $ , and more generally $\mu_{iso}=(C_{ii}-C_{ij}+3C_{ll})/15$ where $i,j=1,2,3; l=4,5,6$, is reported for $\mu$. To agree with the convention used in Materials Project, we also report this value for Ge in table \ref{elastic}.

\subsection{Young modulus}

The Young modulus along $x$ is the response to a uniaxial deformation in that direction:
$Y= \left( \frac{d \sigma_{1}}{d\ep_{1}} \right)_{\si_{j>1}=0} = \frac{1}{S_{11}}$. 

But for isotropic materials, usually $Y_{iso}=9\mu_{iso} B/(3B+\mu_{iso})$, where $B$ and $\mu$ are defined above, is reported.

\subsection{Poisson ratio}

Finally, for a uniaxial deformation along the x axis,  the Poisson ratio is defined as $\nu=-\left( \frac{\partial \ep_{2}}{\partial \ep_{1}} \right) _{\si_{i\ne 1}=0} = -\frac{S_{21}}{S_{11}}$.

Again, for isotropic materials, usually the value $\nu_{iso}=(1.5 B-\mu_{iso})/(3B+\mu_{iso})$ is reported.

As noted, the values of $Y,\mu$ and $\nu$  would depend on the direction of applied stress with respect to crystalline axes in the case of anisotropic materials, where more care needs to be taken when investigating elastic properties theoretically or experimentally.

\subsection{Mode Gruneisen parameters}

The mode Gruneisen parameters are obtained from the phonons and the cubic force constants. They represent the change in the phonon frequencies under applied strain. Even though the strain is usually considered to be isotropic compression or dilation, one can also define the response to an arbitrary strain. By definition, we have $\gamma_{\la,\mu\nu} = - \frac{d ln \,\om_{\la} }{3 d \ep_{\mu\nu}} $. 
But usually the volume derivative, which corresponds to an isotropic strain is considered:
$ \gamma_{\la} = - \frac{d ln \,\om_{\la} }{d ln \, V} $ where $V$ is the unit cell volume. Applying a uniform hydrostatic strain $\epsilon_{\alpha \beta} = \delta_{\alpha \beta} \,dV/3V= \delta_{\alpha \beta} \,d ln \,V/3$. Similar to the group velocity, we  start with the square of the frequency written as the product of the dynamical matrix by eigenvectors: $\om^2=e.D.e$ and consider the change in $D$ under strain: $dD/d\eta= \sum_R d\Phi/d\eta \, e^{ik.R}$, and use $d\Phi/d\eta = \Psi (R+\tau+du^0/d\eta)$. Note the latter term, which is non-zero for low-symmetry structures having their own internal relaxation under strain $u^0(\eta)$, is often neglected in the literature.
In the previous section, we derived this equation $u^0(\eta)=-\Ga (\Pi+Q\eta)$ for $u^0$ in the harmonic approximation. Therefore $du^0/d\eta=-\Ga Q$.
Using the Hellmann-Feynman theorem applied to the derivative of the dynamical matrix, and substituting the above relation for the strain derivative of the harmonic force constants, we finally find:
\begin{multline}
    \ga_{\la}(k)=-\frac{1}{6 \om_{\la}^2(k)} \sum_{\tau,R_1\tau_1,R_2 \tau_2} \frac{ \Psi_{\tau,R_1\tau_1,R_2 \tau_2}^{\al\al_1\al_2} }{\sqrt{ m_{\tau} m_{\tau_1}}}\,  
    e_{\tau \al,\la} (k)^* \, e_{\tau_1 \al_1,\la}(k) \, e^{ik.(R_1+\tau_1-\tau)} \\\times \,\left[(R_2+\tau_2)^{\al_2}-\sum_{\mu=1,2,3} \Ga_{\tau_2 \al_2},. Q_{.,\mu\mu} \right]
    \label{grun}
\end{multline}
The temperature-dependent Gruneisen parameter is defined to be the thermal average of the above modes weighted by their heat capacity: $\ga(T)=\sum_{k\la} c_{k\la}(T) \ga_{\la} (k)/ \sum_{k\la} c_{k\la}(T) $, where the heat capacity per mode is $c_{k\la}(T)= k_B x_{k\la}^2 /{\rm sinh}^2 x_{k\la} $ and $x_{k\la} = \hbar \om_{\la} (k)/2 k_B T$.

\subsection{Thermodynamic properties within the quasi-harmonic approximation (QHA)}

The simplest way to calculate thermodynamic properties at finite temperatures is to use the QHA, which consists of assuming the T-dependence comes only through a volume dependence  caused by thermal expansion. A given property $\cal{P}$ needs only to be calculated as a function of volume, and once the relationship $V(T)$ or the thermal expansion is known, the temperature can be associated to the corresponding volume and the properties calculated as: ${\cal P}^{QHA}(T) \approx {\cal P} (V(T))$. 
To avoid calculation of force constants and phonons at different volumes, we use the definition of the Gruneisen parameter as given in eq.\ref{grun} to obtain phonon frequencies at different volumes. Then the harmonic free energy at a given temperature $T$ is the sum of electronic and vibrational contributions, and calculated using the following formula, which is the sum of zero temperature energy as a function of volume and the vibrational harmonic part:
\begin{equation}
    {\cal F}^{\rm QHA}(T)\approx {\cal F}(V(T))=E_0(V(T)) + k_B T \sum_{\la} {\rm ln} \, [2 \, {\rm sinh} (\hbar \om(V(T)) /2k_BT)] 
\label{qha}
\end{equation}
Here we are neglecting the electronic contribution to the free energy which may become important for metals, and is usually negligible for larger gap materials.
At a fixed temperature $T$, this function can be minimized with respect to the volume to yield the equilibrium volume $V(T)$ to be used in eq. \ref{qha}. 

For the zero-temperature part of the free or total energy $E_0$ one can either assume $E_0(V)=E_0(V_0) + \frac{1}{2V_0} B (V-V_0)^2$, with $B$ being the bulk modulus, or just use the Taylor expansion \ref{vofeta} in powers of $\eta$ and replace volume derivatives by strain derivatives: $d ln V = dV/V_0=3 d\ep$, so that the zero-temperature equation of state can be written as:
\begin{equation}
E_0(V)=E_0(V_0) + \frac{(V-V_0) }{3V_0} \frac{dE_0}{d\ep} +\frac{(V-V_0)^2 }{18 V_0^2} \frac{  d^2 E_0}{d\ep^2} + \frac{(V-V_0)^3 }{162 V_0^3} \frac{  d^3 E_0}{d\ep^3} + ...
\end{equation}
where the strain derivatives are:
\begin{equation}
\frac{dE_0}{d\ep}={\rm Tr}_{\al} \left[ \Xi_{\al\al} - \sum_{\tau\tau'\al'}\Pi_{\tau\al} \,\Ga_{\tau\al;\tau'\al'} \, Q_{\tau'\al';\al\al} \right] \,\, ; \,\,\,
\frac{  d^2 E_0}{d\ep^2} = V_0 \sum_{ij=1,2,3} C_{ij}
\end{equation}
The third derivative can be easily obtained only if $u^0=0$ due to symmetry; otherwise the expression becomes complicated, due to the fact that  equation \ref{V_eff}  has to be extended to third-order and equation \ref{u0} would not be linear in $\eta$ anymore, and needs to be solved for $\eta$ and $\eta$ be substituted back into \ref{V_eff} before taking the derivatives.
Only in the case where $u^0=0$ or $du^0/d\eta=0$, do we have 
\begin{multline}
\frac{d^3 E_0}{d\ep^3}(u^0=0;\frac{du^0}{d\eta}=0)= \\
\frac{1}{6N}\sum_{R_i\tau_i\al_i} \Psi_{R_1\tau_1,R_2\tau_2,R_3\tau_3}^{\al_1\al_2\al_3} (R_1+\tau_1)^{\al_1}(R_2+\tau_2)^{\al_2}(R_3+\tau_3)^{\al_3} 
\end{multline}

In this case, we can also deduce the dimensionless parameter  $$B'_0 =\left(\frac{dB}{dP}\right)_{T=0}=V_0 \left(\frac{d^3 E}{dV^3}\right)/\left(\frac{d^2 E}{dV^2}\right)= \frac{1}{27 V_0 B }\left(\frac{d^3 E_0}{d\ep^3} \right)
$$.

The volume dependence of the phonon frequencies is given by this formula:  $d\om_{\la}/dV=-\gamma_{\la} \om_{\la}/V$ . We can finally find the equilibrium volume $V_{eq}(T)$ by minimizing $F^{\rm QHA}$ with respect to the strain or the volume.

Since the bulk modulus itself is a function of $T$, or, within the QHA, of the volume, we need to consider this dependence in the first term of the free energy when performing volume-minimization. The simplest volume dependence of the bulk modulus is a linear one. So we take $B(V)=B(V_0)(1+B'_0 (V-V_0)/V_0)$ .  In FOCEX, to include a temperature-dependence in the bulk modulus, we therefore use $B(V)\approx B(V_0) (1+3\eta_{\rm eq}(T) B'_0 )$. 
Note that only for materials with no internal relaxations $u^0$ can we easily calculate and use $B'_0$. 

\subsection{Beyond QHA}
If one has access to the mode Gruneisen parameters, the QHA is the simplest way to describe thermal expansion in a crystal. To go beyond QHA, one needs to consider the more accurate anharmonic form of the Free energy. Its volume derivative gives the pressure (mechanical equation of state). Solving the zero pressure equation for the volume then provides the equilibrium volume and thus the thermal expansion. 
Wallace has a discussion of the free energy of an anharmonic system in \cite{Wallace1998}. We also provide a variational formulation of the free energy in \cite{esfarjani2020}. In the latter, the free energy is a sum of an effective harmonic free energy, where the potential energy is  defined by $K$ as in eq. \ref{keff}, and an anharmonic correction as defined by $\langle V^{LD} - K_{ij} y_i y_j/2 \rangle$. This can be a starting point for taking volume or strain derivatives and solve $P(V_{eq})=0$. More on this will be published in our upcoming paper on SCOP8. 
This discussion is limited to insulators when the electronic contribution to thermal expansion is neglected. Allen \cite{allen2022} provides a more extensive discussion of this issue including electronic excitation effects also valid for metals.

\subsection{Optical properties}
The static dielectric constant is the sum of the ion-clamped (electronic) dielectric constant $\ep_{\infty}$ coming from a DFPT calculation, and the dielectric ionic polarizability: $\ep_{0}^{\al\be}(\om)=\ep_{\infty}^{\al\be} + \chi^{\al\be}(\om)$ where:
\begin{equation}
    \chi^{\al\be}(\om) =\frac{1}{4 \pi \epsilon_o} \frac{4 \pi e^2}{V_0} \sum_{\la>4} \frac{{z_{\la}^{\al}}^* z_{\la}^{\be}}{-\om^2 + \om_{\la}^2 - i \eta^2} 
\end{equation}
where $\eta$ is a smearing factor reflecting anharmonicity or disorder, typically on the order of inverse phonon lifetimes, and $z_{\la}^{\al}= \sum_{\tau} Z_{\tau}^{\al\be} e_{\la}^{\tau \be} (k=0) /\sqrt{m_{\tau}}$ is the dipole moment induced by the optical mode $\la$ at the zone center $(k=0)$. Its square $\sum_{\alpha} |z_{\la}^{\al}|^2$ is proportional to the IR peak intensity. This optical ionic response is in the IR domain where the excitation frequency is on the order of phonon vibrational frequencies.

\subsection{Brief algorithm}

1) There are 5 small input files which need to be prepared first: \texttt{default.params, structure.params, dielectric.params, latdyn.params, kpbs.params}.

2) Force-displacement supercell data must be stored in files named \texttt{POSCAR1, POSCAR2, ... ; FORCEDISP1, FORCEDISP2, ...}. The FOCEX code checks consistency between the atomic structures of the supercell and the primitive cell described in  \texttt{structure.params}. Force data may come from supercells of different shapes. Data for  each shape is stored in a different input \texttt{POSCARi, ... ; FORCEDISPi, ...} i=1,2,...  . Typically one large supercell can be used to extract the harmonic FCs, which can be of long range, and a smaller supercell can be used for large-amplitude anharmonic displacements. The files \texttt{POSCARi} contain the equilibrium or reference coordinates of the atoms in the supercell number i, in the POSCAR format defined in VASP.

3) Space group symmetries of the crystal are identified from the primitive cell  information provided in \texttt{structure.params}.

4) The mapping between selected (according to their range) independent FCs and the full set of FCs for each rank is established according to Eqs.(\ref{esfarjani2008method3})  to Eqs.(\ref{space_group}).

5) The  invariance relations, based optionally on translations  (Eqs.(\ref{asr}), rotations (Eqs. (\ref{rotational})) and Huang constraints are coded in a matrix we label by $[ I ]$.
These are a set of linear relations on unknown FCs which will be the homogeneous part of the linear matrix $[ A ]$ equation to solve.

6) Based on displacements in the supercell(s), the second block of the $[ A ]$ matrix, as well as the inhomogeneous part of the array $b$ containing the forces are setup according to Eqs. (\ref{forcemodel}).  So the first block of matrix $[A]$ contains coefficients reflecting invariances and the second block contains first or higher powers of displacements.

7) The linear, over-determined system $A \Phi=b$ is solved using a singular-value decomposition (SVD) or ridge regression algorithm. The unknown array $\Phi$ contains the selected independent force constants.  

8) Results are written into several files that can be used for testing and further analysis.

9) Finally, using the harmonic FCs, the phonon dispersion, density of states (dos), Gruneisen parameters, elastic constants and some thermodynamic properties based on QHA and dielectric constant are calculated and written into files.

\begin{figure}[h]
    \centering
    \makebox[\textwidth][c]{\includegraphics[width=0.73\textwidth]{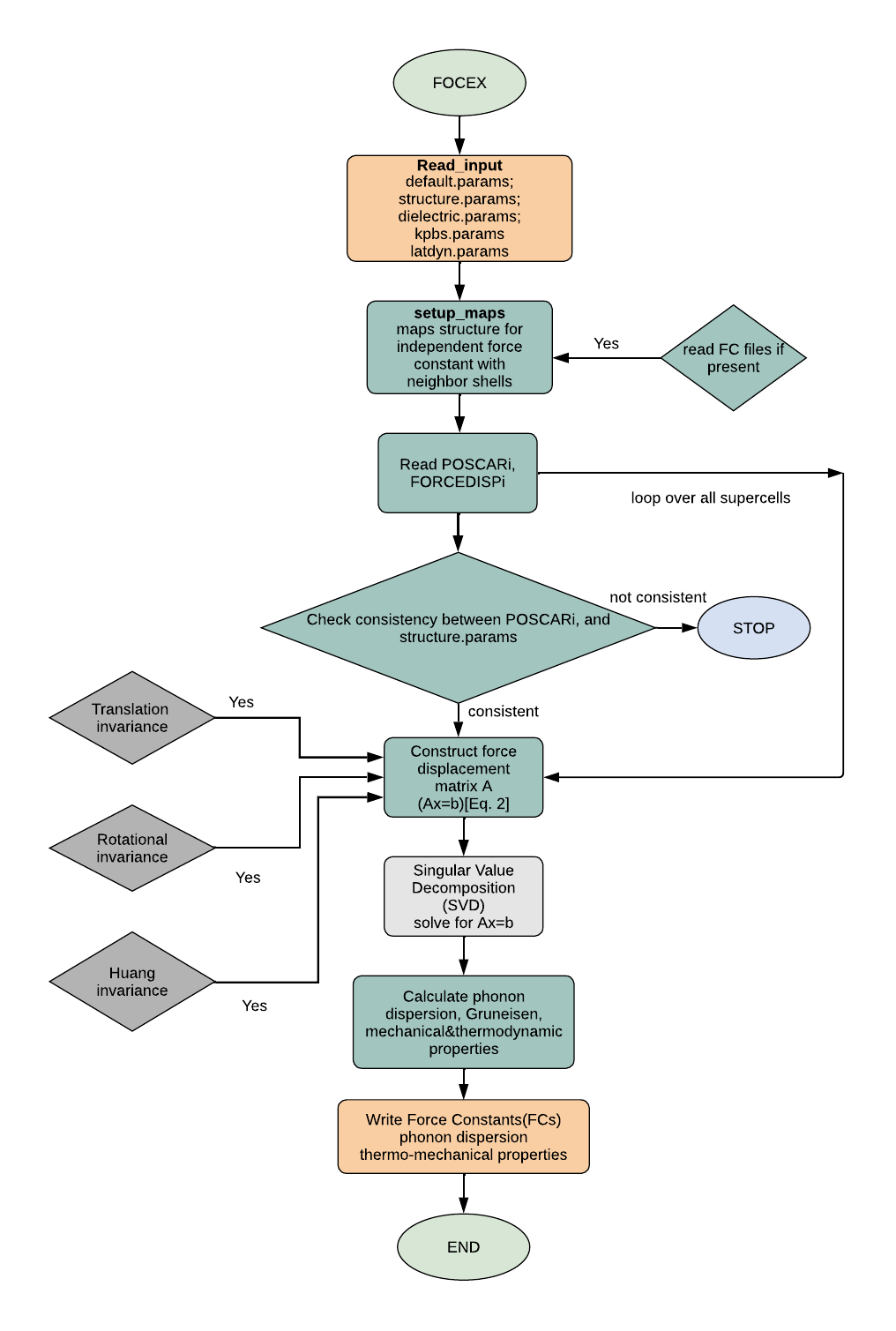}}
    \caption{Workflow of FOCEX}
    \label{focex_flowchart}
\end{figure}

\subsection{Input files} 
The following files need to be present in the folder where the code FOCEX is run:


0) \texttt{default.params} file contains some of the less-used parameters which  can be initialized if their default values do not work. This file needs to be present with lines initialized by 0 (in case the default has to be used), otherwise the defaults can be overwritten by whatever number appears on that line. The first line is the tolerance to equate two coordinates (it is 0.002 Ang), the second is the regularization parameter in the ridge regression and is set to 1E-10. The third is the number of translation vectors defining the outer shell of neighbors. It is set to 7, and can be increased if the supercell is larger than 12 translation vectors.
The user does not need to touch this file, but it needs to be present in the working directory.

1) The unit cell information and type and range of force constants to be fitted are specified in the file \texttt{structure.params}. These include conventional and  primitive lattice vectors, atom identities: (name, mass, and positions in reduced units of the conventional cell), number of neighbors to include for each rank, options for symmetries to impose (translational, rotational and Huang), options for rank of force constants (2,3,4, ... up to 8) to be included in the fitting.

2) The dielectric matrix  $\epsilon_{\infty}$ and Born charges $Z_{\tau}$, in the same order as they appear in the \texttt{structure.params} file are read from the \texttt{dielectric.params} file. The dielectric matrix and Born charges are obtained from a separate DFPT calculation. 

3) The information regarding the reference supercells are read from files \texttt{POSCAR1, POSCAR2, ...} written in the same format as VASP POSCAR file. The corresponding cartesian displacements (in Ang) and forces (in eV/Ang) are read from files \texttt{FORCEDISP1, FORCEDISP2, ...}. Each \texttt{FORCEDISPi} file contains all the force-displacement snapshots corresponding to the $i^{th}$ supercell defined in  \texttt{POSCARi}. 
Each snapshot is separated from the next one by two lines, the first of which should contain the word "POSITION" and the second one contains the total energy of that snapshot. The N following lines contain the cartesian coordinates of the N atoms followed by the forces acting on them (6 columns in total).

4) The file \texttt{kpbs.params} contains the list of k-points to use for the phonon band structure calculations. After the force constants are extracted, the FOCEX code also computes the harmonic properties, namely the phonon band structure, elastic and thermodynamic properties. 

5) The input file \texttt{latdyn.params} contains information needed for the thermodynamic properties, namely the size of the k-point mesh in the first-Brillouin zone, the frequency mesh for the density of states, the smearing factor, and finally the temperature window for the thermodynamic properties.


\subsection{Output files} 

Many output files are generated after the run. Many of them are used to test the validity of the output in case something goes wrong. The main useful files are:

1) \texttt{fc1\_irr.dat, fc2\_irr.dat, fc3\_irr.dat, fc4\_irr.dat, ...}  contain the irreducible FCs (selected based on crystal symmetries and imposed range). 

2) \texttt{fc1.dat, fc2.dat, fc3.dat, fc4.dat, ...} contain the full set of FCs, in  separate files  for ranks 1, 2, 3, 4 ... 
Each line contains the atom labels $R_i\tau_i$ involved, the cartesian coordinates $\alpha_i$ and the FC value $\frac{\partial^n V}{ \partial u_{R_1\tau_1\alpha_1} \partial u_{R_2\tau_2\alpha_2} ...} $. 

3) \texttt{lat\_fc.dat} contains the information on the primitive cell and neighboring cells, as well as the number of independent and dependent force constants.

4) \texttt{log-<extra>.dat} contains the log of the run, where the parameters "<extra>" label the type of the run, as described in the file \texttt{ios.f90}.

5) \texttt{mech.dat} contains the mechanical properties (elastic constants, etc...). 

6) \texttt{thermo\_QHA.dat} contains the thermodynamic properties: temperature, total energy, free energy, heat capacities, entropy, equilibrium QHA-volume, thermal expansion coefficient and average Gruneisen parameter. 

7) \texttt{temp\_free.dat} contains the free energy, pressure, Gruneisen parameter and bulk modulus in the $(V,T)$ plane.

8) \texttt{bs\_grun.dat} and \texttt{bs\_freq.dat} both contain the band structure. While the first also has the Gruneisen parameters, the second contains phonon group velocities. The same data in the irreducible Brillouin zone is written in \texttt{ibz\_bands.dat} with group velocities, \texttt{ibz\_grun.dat} with Gruneisen parameters. 

9) \texttt{ibz\_dos.dat} contains the Gaussian-smeared DOS from k-points generated in the IBZ, while \texttt{fbz\_dos.dat} contains the DOS generated using the tetrahedron method.

10) \texttt{chi\_real.dat} and \texttt{chi\_imag.dat} contain the 9 matrix elements of the real and imaginary parts of the ionic polarizability versus frequency.

11) Other auxiliary files contain information on the matrix $[ A ]$ (\texttt{amatrx.dat}) and the data generated during the SVD procedure (\texttt{svd-results.dat}), and many other files used for book-keeping and debugging, not necessarily of use to the user.

\subsection{Utilities} 

Some utilities (\texttt{read\_outcar.f90}  and \texttt{read\_qe.f90})  are also provided to convert the log or output files from a single snapshot of VASP (OUTCAR)   or Quantum Espresso to FORCEDISP format. A shell script \texttt{xtract.sh} then calls these executables, and appends the output from each snapshot into the file \texttt{FORCEDISP1}. In addition, files for plotting the outputs using the Gnuplot software are also provided. \texttt{all.plt} or \texttt{bvel.plt} plot the phonon band structure,  Gruneisen parameters DOS, and the group velocities.  \texttt{KTICS.BS} should be imported in this file. \texttt{fcs.plt} plots the harmonic force constants versus pair distance and visualizes them in the Wigner-Seitz cell associated with the supercell. \texttt{qha.plt} plots the results from the QHA calculations: total energy, free energy, strain, volume, heat capacity, entropy and bulk modulus versus temperature. \texttt{kp.plt} visualizes the full zone, irreducible zone and band structure k-points in 3D reciprocal space. 

\section{ANFOMOD}
The code ANFOMOD performs molecular dynamics simulations with the anharmonic force field developed by FOCEX. It takes as input the output files of FOCEX: \texttt{lat\_fc.dat, fc1.dat, fc2.dat, fc3.dat, fc4.dat}. It also has its own control file \texttt{anh\_md.params} in which the type of run (NVE or NVT), duration, time step, temperature, .... are specified. Finally, supercell information and Cartesian atomic coordinates are read from \texttt{coordinates.sc} in VASP POSCAR format.

Using the polynomial force field, assuming the residual force $-\Pi$ is zero, the force acting on an atom $R\tau$ is given by:


\begin{eqnarray}
  {F_{R\tau}^{\alpha}}^{LD}=-\sum_{R_1\tau_1} \Phi_{R\tau,R_1\tau_1}^{\alpha\beta} u_{R_1 \tau_1}^{\beta} -\sum_{R_1\tau_1 ,R_2\tau_2}\frac{1}{2}   \Psi_{R\tau,R_1\tau_1,R_2\tau_2}^{\alpha\beta\gamma} u_{R_1 \tau_1}^{\beta}u_{R_2 \tau_2}^{\gamma} \nonumber \\
  -\sum_{R_1\tau_1,R_2\tau_2,R_3\tau_3}\frac{1}{6}   \chi_{R\tau,R_1\tau_1,R_2\tau_2,R_3\tau_3}^{\alpha\beta\gamma\delta} u_{R_1 \tau_1}^{\beta} u_{R_2 \tau_2}^{\gamma} u_{R_3 \tau_3}^{\delta}
\end{eqnarray}

\vspace{\columnsep}

To insure translational invariance, in the actual code the quantity $u_{R\tau}$ has been subtracted from every single displacement, i.e. $u_{R_1 \tau_1}^{\beta}$ is replaced by $u_{R_1 \tau_1}^{\beta}-u_{R \tau}^{\beta}$, and so on.

To update atomic positions, the standard velocity Verlet \cite{Ohno1999} algorithm is used. If canonical ensemble needs to be simulated, a Langevin thermostat, in which the damping rate can be adjusted, has been adopted. A good initial guess for the latter would be $1/(1000 dt)$ where $dt$ is the MD time step. 

The heat current is formulated from the following definition: $J^{\alpha}=\sum_{R\tau} j_{R\tau}^{\alpha}$
where $R$ runs over the primitive cells in the simulation supercell, and where the local heat current is defined as:

\begin{equation}
    j_{R\tau}^\alpha=\frac{1}{4}\sum_{R_1 \tau_1} (R_1+\tau_1-R-\tau)^{\alpha} 
    f_{R\tau,R_1\tau_1}^{\beta}   (v_{R_1 \tau_1}+v_{R \tau})^{\beta}
\label{localjq}
\end{equation}

where $f_{R\tau,R_1\tau_1}^{\beta} =\frac{ \partial \V_{R\tau}}{ \partial u_{R_1\tau_1}^{\beta}} $ is the effective pair force between $R\tau$ and $R_1\tau_1$.  $\V_{R\tau}$ is the local potential energy of site $R\tau$:
\begin{equation} 
\begin{split}
 \V_{R\tau}=\sum_{\alpha} u_{R \tau}^{\alpha} \Big[ \sum_{R_1 \tau_1} \frac{1}{2}   
 \Phi_{R\tau,R_1\tau_1}^{\alpha\beta} u_{R_1 \tau_1}^{\beta} +\sum_{R_1\tau_1 ,R_2\tau_2} \frac{1}{3!} \Psi_{R\tau,R_1\tau_1,R_2\tau_2}^{\alpha\beta\gamma} u_{R_1 \tau_1}^{\beta}u_{R_2 \tau_2}^{\gamma} \\
  +\sum_{R_1\tau_1,R_2\tau_2,R_3\tau_3} \frac{1}{4!} \chi_{R\tau,R_1\tau_1,R_2\tau_2,R_3\tau_3}^{\alpha\beta\gamma\delta} u_{R_1 \tau_1}^{\beta} u_{R_2 \tau_2}^{\gamma} u_{R_3 \tau_3}^{\delta} \Big]
\end{split}
\end{equation}

The total, kinetic and potential energy as well as the heat current will be written in separate files at a given output-rate. To obtain the thermal conductivity using the Green-Kubo formula, the heat current data needs to be post-processed. 

\section{Other codes}

Other codes have been developed which will read the FOCEX outputs, namely \texttt{fci.dat} (i=2,3,4,...) and \texttt{lat\_fc.dat} to compute other physical properties.  Details about these codes will be published elsewhere.

\begin{itemize}
    \item {\bf THERMACOND} computes the thermal conductivity as the solution to the Boltzmann transport equation within and beyond the relaxation time approximation (RTA).

    \item{\bf SCOP8} calculates for a given temperature, the state of equilibrium of the crystal, namely its primitive cell shape and size as well as the atomic coordinates, and the renormalized force constants calculated within the self-consistent phonon (SCP) approximation.

\end{itemize}

\section{Installation} 

The alatdyn software package is available on GitHub under GPL-3.0 license  at \url{https://github.com/KeivanS/Anharmonic-lattice-dynamics}, where each code is stored in the corresponding subdirectory. A \texttt{Makefile} template for each code has been provided. After cloning the repository, the user can type \texttt{make} to generate all needed executables, or alternatively go into each subdirectory, edit the \texttt{Makefile} according to their system and needs, and run \texttt{make}. The codes can be compiled using the GNU Fortran Compiler \texttt{gfortran} or the Intel Fortran Compiler \texttt{ifort}, and are tested to be compatible with \texttt{gfortran} versions 7-11 and \texttt{ifort} versions 18-19. (Newer and/or older versions are likely to  be compatible as well.) The only external library is the zhegv diagonalization routine from LAPACK. A successful compilation results in the creation of the \texttt{alatdyn\_BIN} directory containing all the executables. 

In a typical high-performance computing cluster environment, one should be able to compile the codes using commands similar to the following. For example, to compile FOCEX with GCC (assuming gfortran is the install Fortran compiler):
\begin{verbatim}
module load gcc
git clone https://github.com/KeivanS/Anharmonic-lattice-dynamics
cd Anharmonic-lattice-dynamics/FOCEX
make
\end{verbatim}
assuming GCC is preinstalled as a module on the cluster.

\section{Example of Ge, and some tips}
Here we show the example of Germanium as a system ran with FOCEX. We must emphasize that the force-displacements inputs stored in \texttt{FORCEDISP1} are up to the user. For a choice of the supercell and magnitude of displacements, the best set of FCs will be extracted. But the results will change with a change in the supercell, magnitude of displacements, choice of the exchange-correlation potential, lattice parameter, chosen rank and chosen range of FCs. So it is up to the user to make sure the obtained results are converged. As a rule of thumb, a supercell of 100 atoms or more is usually enough to provide a very reasonable set of force constants. To get accurate harmonic force constants, displacements need to be small enough. Typically displacements on the order of or less than 0.05 \AA $\,$ are reasonable. While many codes require moving one or two atoms at the most, in FOCEX all atoms maybe simultaneously moved, providing more information on FCs for the same data size. A reasonable scheme would also be to move all atoms as a superposition of the normal modes sampled from the canonical ensemble, if effective force constants at higher temperatures need to the be extracted. If normal modes are not known, as is usually the case, one can still adopt normal modes generated with an arbitrary but cheap force field in order to avoid two separate supercell DFT calculations. The rationale for this approach is that direction of normal modes is defined by symmetry and the environment of the atom and is less dependent on the details of the force field. Generation of such displacements sampled from the canonical ensemble is done in the utility file \texttt{sc\_snaps.x} which generates supercell snapshots of a required number if a temperature of interest is given. In principle, one or two snapshots in which all atoms are moved maybe sufficient to generate harmonic force constants. If only harmonic FCs are required, it is best to generate snapshots in which only inequivalent atoms are moved. More snapshots will be needed if anharmonic FCs are also required, typically on the order of 20, depending on their rank and chosen range. The number of generated FCs is a polynomial function of the range and an exponential function of the rank. Therefore more data (snapshots) would be needed to solve for all the unknowns if either range or rank is increased. The FOCEX code prints out for a chosen rank and range how many unknown FCs will be generated and therefore at least as many equations on forces vs displacements (snapshots) need to be generated.  Checks must be performed to make sure the generated FCs don't change when the range is increased; just the furthest ones are added and are smaller than the existing ones. Another check for convergence is that the physical quantities of interest such as the phonon dispersion and the Gruneisen parameters do not appreciably change as ranges are increased. Largest (infinite norm) and average (first norm) error, as well as the percentage deviation (second norm) of the fitted forces  are written at the end of the log file.

\begin{figure}[h]
    \centering
    \makebox[\textwidth][c]{\includegraphics[width=0.89\textwidth]{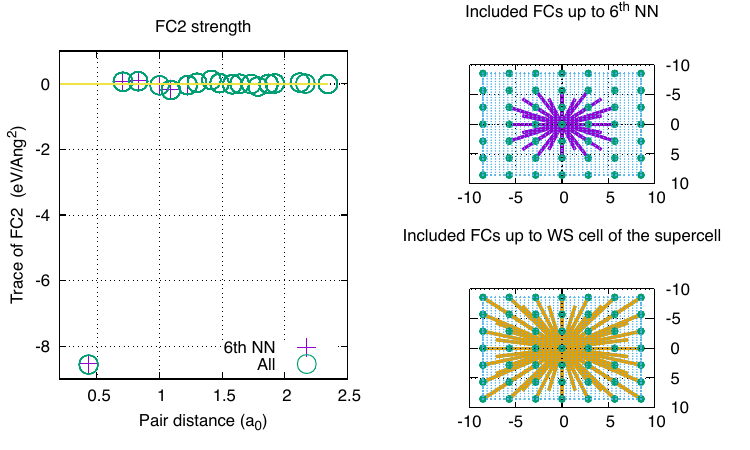}}
    \caption{Output of the fcs.plt: On the left are the fitted Ge harmonic force constants in units of the conventional lattice parameter $a_0=5.7 \AA$, when only up to 6th nearest neighbors are included in the fitting (plus sign) and when all pairs within the supercell are included (open circles). Right: To visualize included force constants, we show the Wigner-Seitz (WS) cell of the supercell (in blue) from the top, centered on an atom, the grid of primitive translation vectors (green spheres), and selected force constants (lines) connecting the central atom to its neighbors within the supercell. Top right: only up to 6th neighbors, bottom right: all FCs within the supercell. This can be visualized for every atom in the primitive cell, but for Ge both atoms are equivalent. }
    \label{fcs}
\end{figure}

\begin{figure}[h]
    \centering
    \makebox[\textwidth][c]{\includegraphics[width=0.99\textwidth]{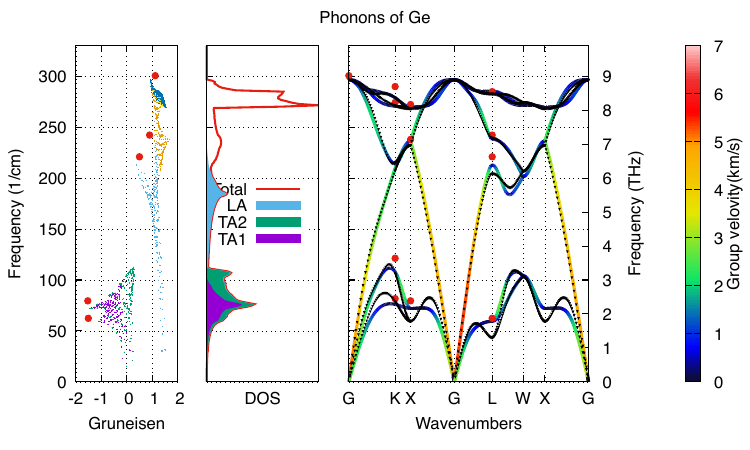}}
    \caption{Mode Gruneisen parameters in the irreducible Brillouin zone (left), phonon density of states (center), and dispersion of Ge (right) calculated by FOCEX. Large red dots are experimental data. Small dots in black show the phonon dispersion curve as a result of including only 6 nearest neighbors in the fitting. TA modes are very sensitive to the number of neighbors. Near the X point TA dispersion should be flat. This can be achieved if at least 9 nearest neighbors are included. Color bar shows the group velocity magnitude in km/s}
    \label{gedispersion}
\end{figure}

For Ge, we have used a supercell of 216  atoms with 36  snapshots in which only one or two atoms have been moved. Forces were calculated using VASP\cite{vasp1,vasp2,vasp3,vasp4} with the GGA-PBEsol functional, at cutoff energy of 225 eV and a Gamma-centered 7x7x7 k-point mesh within the 3x3x3 supercell. The lattice constant of $a=5.702 $ \AA  was found using 14x14x14 k-mesh for the primitive cell optimization calculations. This is only 0.7\% larger than the experimental value of $5.66$ \AA. 
All harmonic FCs within the supercell were included in the fitting as shown in Fig. \ref{fcs}. Cubic force constants were included up to the 6th nearest neighbors.
The results on the phonon dispersion and Gruneisen parameters are displayed in Fig. \ref{gedispersion}. The slight discrepancy between the computed and experimentally measured\cite{nilsson} frequencies is most likely due to the larger PBE lattice constant of 5.7 \AA, which is slightly larger than the experimental value of 5.66 \AA. This is a 2.1\% larger volume. Therefore for a Gruneisen parameter of +1 for instance, a 2.1\% reduction in the phonon frequency should be expected. The Gruneisen parameters in Fig.\ref{gedispersion} show a very decent agreement with the values of Olego and Cardona \cite{cardona}.

To view the set of harmonic force constants used in the fitting, one can make a plot similar to Fig.\ref{fcs} by using the file \texttt{fcs.plt}.

The dispersion compares well with other available data. The elastic tensor is summarized in table \ref{elastic}. There is usually some fluctuation in the available data, depending on the temperature where measurements were taken, and on the numerical side, results depend on the lattice parameter and exchange-correlation functional used, and the method used to extract elastic constants. 

For a crystal of cubic symmetry, the bulk $B$, Young $Y$ and shear $\mu$  moduli are respectively given by: $B=(C_{11}+2C_{12})/3 ; Y=1/S_{11}=3B (C_{11}-C_{12})/(C_{11}+C_{12}) ; \mu= C_{44}; \nu=-S_{12}/S_{11}=C_{12}/(C_{11}+C_{12})$. Note the expression for $Y$  and $\nu$ assume deformation along the [100] direction, and this is what we report in the table \ref{elastic} below, in addition to their ``isotropic'' values, as defined in the previous section.

\begin{table}[h]
\centering
\begin{tabular}{c||c|c|c|c|c|c|c|c|c|c|}

&$C_{11}$ & $C_{12}$ &$C_{44}$ & B & $Y_{100}$ & $Y_{iso}$ & $\mu_{100}$ & $\mu_{iso}$ & $\nu_{100} $ & $\nu_{iso}$  \vspace{0.8 mm} \\\hline \hline
LD model & 121 & 69 & 54.0 &  86.4 & 73 & 105 & 54 & 41  & 0.36 & 0.29  \\ \hline
Ioffe   & 126 & 44 & 67.7 & 71.3 & & 103 & & 41 & & 0.26 \\ \hline
McSkimin & 129 & 48.3 & 67.25 & 77 &  102.5& & 67.1 & & & 0.26 \\  \hline
MP (DFT) & 102 & 36 & 56 &  58 & 83.3 & & & 47 & & 0.25$^*$ \\ \hline
\end{tabular}
\caption{Elastic constants of Ge crystal in GPa. Experimental data from McSkimin\cite{mcskimin}  is at T=298K, and those of Ioffe Institute \cite{IoffeGe} are at 300K. The reported value of 0.19 on MP is not consistent with $-S_{12}/S_{11}=0.25$, but we are reporting the latter. For the Young and shear moduli and the Poisson ratio, we are reporting them for both deformations along [100] and the average value denoted by the subscript {\it iso}.}
\label{elastic}
\end{table}

Group velocities are also compared to the data in Ioffe site\cite{IoffeGe}. Results, which are presented in table \ref{vgroup}, show  relatively good agreement. It is worth noting that there is a simple relationship between group velocities near the zone center (sound speeds) and elastic constants in a cubic crystal. All 3 elastic constants can be extracted from the 3 speeds of sound along [110]: 
\begin{multline}
       \rho \, v_L^2 (100) = C_{11} \, \,\rho \, v_{T1}^2 (100)=  \, \rho \, v_{T2}^2 (100) = C_{44}  \nonumber \\
    \rho \, v_L^2 (111) = \left(\frac{C_{11}+2C_{12}+4C_{44}}{3}\right) \,\,  \rho \, v_{T1}^2 (111)= \rho \, v_{T2}^2 (111) = \left(\frac{C_{11}-C_{12}+C_{44}}{3}\right)   \nonumber  \\
        \rho \, v_L^2 (110) = \left(\frac{C_{11}+C_{12}}{2}+C_{44}\right) \,\, \rho \, v_{T1}^2 (110) = C_{44}    \,\,  \rho \, v_{T2}^2 (110) = \left(\frac{C_{11}-C_{12}}{2}\right)  \nonumber 
\end{multline}

\begin{table}[h]
\centering
\begin{tabular}{c||c|c|c|}

Direction & $ [100]$ & $[110]$ & $[111]$   \\ \hline \hline
This work (LD model) & 3.44 4.76  & 2.77 3.78 5.11  & 3.26 5.43  \\ \hline
Experiment (Ioffe) & 3.57  4.87  & 2.77 3.57 5.36  & 3.06 5.51 \\
 \hline
Experiment (McSkimin) & 3.54  4.91  & 2.75 3.55 5.41  & - -  \\
 \hline
\end{tabular}
\caption{Group velocities (km/s) of the three acoustic branches compared to experimental data from McSkimin\cite{mcskimin} and Ioffe Institute\cite{IoffeGe} taken at room temperature. The largest number corresponds to the longitudinal mode. Along [100] and [111] the two transverse acoustic branches are degenerate.}
\label{vgroup}
\end{table}

\begin{figure}[h]
    \centering
    \makebox[\textwidth][c]{\includegraphics[width=0.99\textwidth]{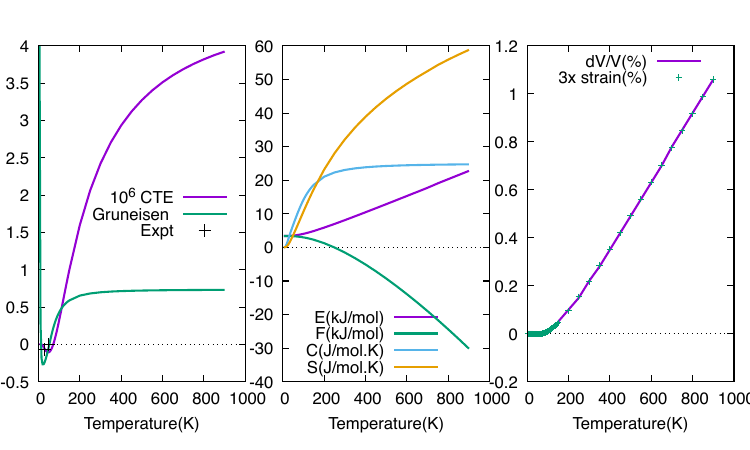}}
    \caption{Left: Gruneisen parameter and linear thermal expansion versus T. Center: Total energy, free energy, specific heat and entropy as a function of T. Right: strain or volume expansion versus T. }
    \label{qha}
\end{figure}

Thermodynamic properties within QHA can also easily be computed. This requires the knowledge of Gruneisen parameters in order to predict phonon dispersion at different volumes. To find the equilibrium volume at a given temperature, the harmonic free energy is then minimized with respect to the volume.  
Figure \ref{qha} shows the plots of thermodynamic potential and its temperature derivatives, as well as the Gruneisen parameter and the linear thermal expansion. Although the zero CTE point at 50K \cite{IoffeGe} is correctly predicted, the CTE value is underestimated by QHA. Experimental data is $6 10^{-6}$ at 500K. There is also a measurement of CTE by Sparks\cite{sparks}, which shows a minimum at 30 K. These two low-temperature data points are shown in Fig. \ref{qha}.

Finally, we must add that this code has been used many times and its results validated in the past 15 years or so. The results on force constants or phonons and lattice thermal conductivity have been published as follows: Silicon\cite{esfarjani-prb2011,minnich-prl2011,Tian2011,scott2018}, half-heuslers\cite{Shiomi2011}
FeSb2\cite{Liao2014}, graphene\cite{Mingo2010,Lee2015},
GaAs\cite{Maznev2013}, PbSe and PbTe\cite{Shiga2012,Tian2012a},
Bi and Sb\cite{Lee2015}.

\section*{Code availability}
The Anharmonic LAttice DYNamics (alatdyn) suites of codes may be accessed and downloaded from \\
https://github.com/KeivanS/Anharmonic-lattice-dynamics.

Documentation is available at: \\
https://alatdyn.readthedocs.io/en/latest/index.html

\section*{Data availability}
the data presented on Germanium, including inputs to FOCEX and all the generated output files can be accessed at:
https://doi.org/10.18130/V3/MVKFQ7

\section{Acknowledgements}
Financial support of SNS, YL, BT, RS and KE by NSF-CSSI award \#2103989 are gratefully acknowledged.

\bibliographystyle{elsarticle-num}

\bibliography{aladyn-refs}

\end{document}